\documentclass[a4paper,11pt]{article}
\pdfoutput=1 

\usepackage{jinstpub} 

\usepackage{lineno}

\title{\boldmath Efficient noble gas purification using hot getters and gas circulation by convection}

\author[a,1]{J.M.R. Teixeira,\note{Corresponding author.}}
\author[a]{C.A.O. Henriques,}
\author[a]{P.A.O.C. Silva, }
\author[a]{R.D.P. Mano,}
\author[a]{J.M.F. Dos Santos,}
\author[a, 1]{and C.M.B. Monteiro}

\affiliation[a]{LIBPhys, Physics Department, University of Coimbra, \\ P-3004-516 Coimbra, Portugal}

\emailAdd{jmrt@uc.pt, cristinam@uc.pt}

\abstract{Noble gas radiation detectors with optical readout are gaining popularity in fields like astrophysics and particle physics due to their ability to produce both ionization and scintillation signals in response to ionizing radiation interaction. In addition, the amplification of primary ionization signals can be achieved by promoting secondary scintillation in the gas.
Noble gas purity, especially concerning impurities like H$_{2}$O, N$_{2}$, O$_{2}$, CO$_{2}$, and hydrocarbons, greatly influences its performance. These impurities can cause the loss of primary electrons and quench the scintillation signal. A very high purity level of the gas is required.
In the early 90's, a simple method was developed for noble gas purification in sealed, small volume (up to few litters) gas radiation detectors. Gas purification is achieved promoting gas circulation through Zr-based hot getters, simply maintaining the gas circulation by convection. The effectiveness of this method has been only confirmed by the energy resolutions achieved in those detectors, which were similar to that achieved in other high-performance noble gas detectors.
In this work, we used waveform analysis of the primary and secondary scintillation signals and we were able to evaluate the impact of the attachment and quenching caused by impurities in one of our detectors filled with pure Xe, and estimate upper values for the impurity content in the gas. The maximum overall impurity concentration was estimated to be below 4 ppm, considering nearly all the impurities, and below 82 ppm if N$_{2}$ is considered. The electron lifetime was measured to be 2.1 $\pm$ 0.1 ms, in line with those achieved in other high-performance optical detectors. 
}


\begin{document}

\maketitle

\flushbottom

\section{Introduction}
\label{sec:intro}

Noble gas radiation detectors with optical readout are being increasingly applied in fields like astrophysics, particle physics and x- and gamma ray imaging. They are based on the notable characteristic of responding to ionizing radiation interactions by producing both primary ionization and scintillation signals. In addition, another important asset of noble gases is the possibility to obtain the amplification of the primary ionization signals by promoting secondary scintillation in the gas, upon drifting the primary ionization electrons throughout the gas into a specific region where gas electroluminescence (EL) takes place.

One of the well-known factors that strongly influences the detector performance is the composition of the gas and its  impurity content (e.g. H$_{2}$O, N$_{2}$, O$_{2}$, CO$_{2}$ and hydrocarbons). The extremely large number of collisions that drifting electrons undergo with the gas atoms/molecules along their drift path may result in the loss of primary electrons to electronegative impurities, due to attachment, or in the loss of the electron energy through inelastic collisions with molecular impurities, due to excitation of rotational or vibrational states of the molecules, thus reducing the scintillation output. In addition, the quenching of the noble gas excited atoms and/or excimers by the molecular impurities also results in reduced electroluminescence output. All these effects contribute to the deterioration of the gas scintillation [1,2] and imposes a stringent purity level of the gas. This is particularly significant if gas EL is achieved in the so-called proportional regime, where the electric fields are below the gas charge multiplication threshold \cite{henriques_1, henriques_2}. 

In general, the detector volume is kept in sealed mode, being the gas purity achieved and maintained by promoting the gas circulation through non-evaporable getters.

Since 1990, in Coimbra, we developed a simple method for gas purification in our gas noble gas chambers, radiation detectors, operated in sealed mode \cite{dos_santos}. We chose non-evaporable getters having a relatively low activation, as low as 350$^{\circ}$C \cite{saes}, and operation temperatures as low as 100$^{\circ}$C. 
The getters fill a vertical tube inserted in the gas circulation loop, figure \ref{fig:gpsc_getters}. Heating tapes are rolled up around the tube on the outside, to activate and to maintain the getters at suitable temperatures inside the tube without the need of feedthroughs to heat the getters from inside, being the gas circulation promoted by convection. Washer-shape getter pieces have been chosen to increase the getter surface exposed to the gas, while facilitating gas convection speed. We have been applying this purification technique to gas chambers of several tenths of litres to few liters in volume, without resorting to ultra-high vacuum techniques to build the chamber, using instead low vapour-pressure epoxies for ceramic- and polyimide-to-metal seals and without resorting to bake out. \cite{dos_santos}.

Up to now, the effectiveness of the gas purification has been confirmed by the good performance achieved in our detectors: (i) The state-of-art energy resolution obtained for 5.9 keV x-ray absorption in Xe and (ii) the fast scintillation output recovery, when the gas circulation is resumed after being stopped for some time and the scintillation being reduced due to impurity increase inside the detector. Nevertheless, through the years, our peers have questioned about the actual effectiveness of our purification method. Besides the above facts, we did not evaluate the real impact of our gas purification technique either in terms of impurity content or in terms of electron attachment or scintillation quenching by impurities. A simple and easy determination of  the impurity content by means of mass spectrometry has been difficult since we are dealing with very low impurity concentrations and large backgrounds at the Residual Gas Analyser (RGA) chamber.

In this work, we used waveform analysis of the primary and secondary scintillation pulses in a gas proportional scintillation counter and we were able to evaluate the impact of the attachment and quenching by impurities in pure Xe and estimate upper values for the impurity content in the gas.

\section{Experimental Setup}
\label{sec:exp_setup}
\begin{figure}[ht]
\centering 
\includegraphics[width=0.8\textwidth,clip]{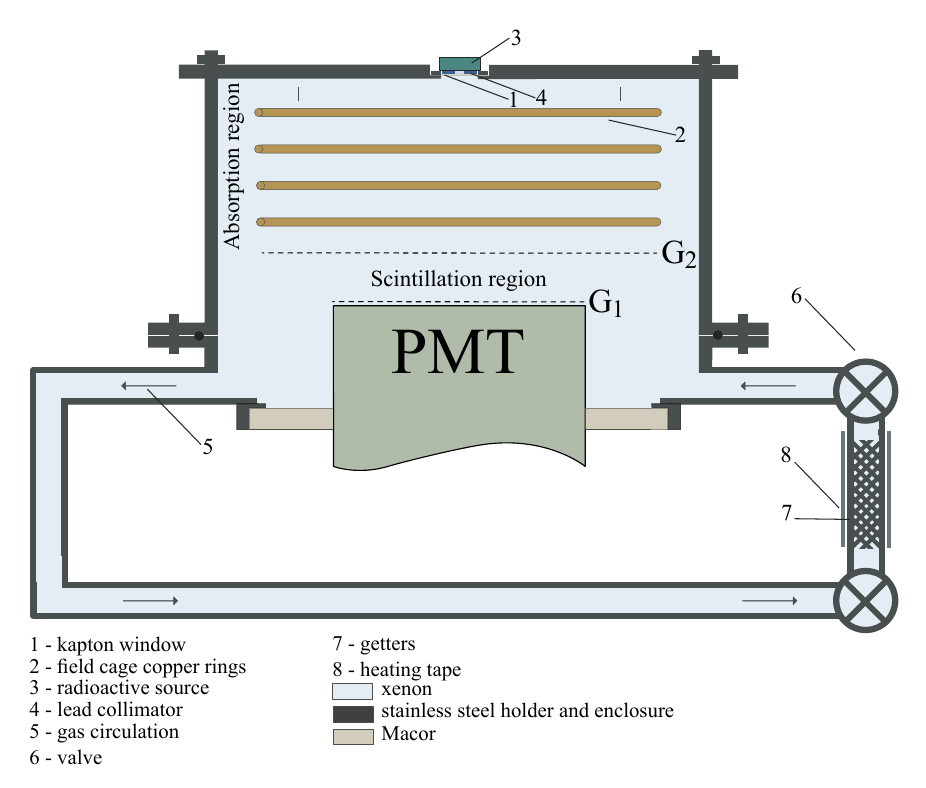}
\caption{\label{fig:gpsc_getters} Schematic of the GPSC used in this work and its gas purification system.}
\end{figure}

For detailed studies of the efficiency of our gas purity technique under well-controlled conditions, we employ a Gas Proportional Scintillation Counter (GPSC), figure \ref{fig:gpsc_getters}, already described in detail in a previous study \cite{henriques_3}. The X-ray interactions in the absorption region lead to the production of a primary scintillation pulse and to the formation of a primary electron cloud. The applied electric field in this region is only strong enough to prevent electron-ion pair recombination, remaining below the gas EL threshold. Under this field, the electron cloud drifts to the EL region, where the electric field intensity is set between the gas EL and ionization thresholds. In this way, electrons excite but do not ionize the Xe atoms, producing a secondary scintillation pulse through the EL process, as the Xe atoms de-excite, being the scintillation readout by the PMT. The PMT output was connected directly to a WaveRunner 610Zi oscilloscope from LeCroy, with a 10 GS s$^{-1}$  sampling rate, directly recording the scintillation signals from the PMT output.

\begin{figure}[ht]
\centering 
\includegraphics[width=0.8\textwidth,clip]{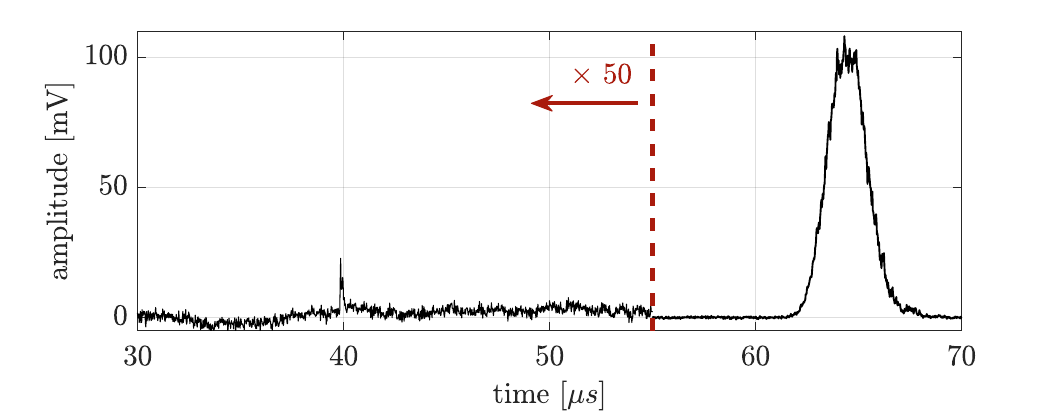}
\caption{\label{fig:single_photon} Typical PMT waveform for 22.1-keV x-rays showing both the primary (S1) and the secondary (S2) scintillation signals.}
\end{figure}

As both primary and secondary scintillation signals, S1 and S2, respectively, are crucial for our study, we optimized the setup to measure them simultaneously with high amplitude resolution. We split the PMT signal into two oscilloscope channels: a "full-channel" to record the S2 pulse without saturation, and a "zoomed-channel" optimized for S1 measurements. The oscilloscope is triggered using the rising edge of the S2 pulse on the "full-channel". A background discrimination algorithm was developed for pre-processing the data. This algorithm rejects background events and waveforms unsuitable for further analysis, based on their duration, time offset, shape, and baseline cleanliness. In addition, every waveform is corrected for baseline offset. The detailed waveform analysis is described in \cite{henriques_3}. Figure \ref{fig:single_photon} depicts a typical waveform obtained for a 22.1 keV x-rays. 

Neutral bremsstrahlung emission produced by the scattering of drifting electrons on Xe atoms \cite{prx} is negligible and do not impact on both S1 and S2 pulses. 

To accurately calibrate the PMT gain, we used a blue LED biased by a direct current to measure the PMT single-photoelectron charge distribution. A mean charge value per single photoelectron of 10.9 +/- 0.3 mV ns was obtained for a PMT biasing of 1450 V. As the PMT used in this work is old, we also took into account in its calibration afterpulsing effects.

\section{Experimental results and discussion}
\label{sec:experimental}

\begin{figure}[ht]
\centering 
\includegraphics[width=0.8\textwidth,clip]{ 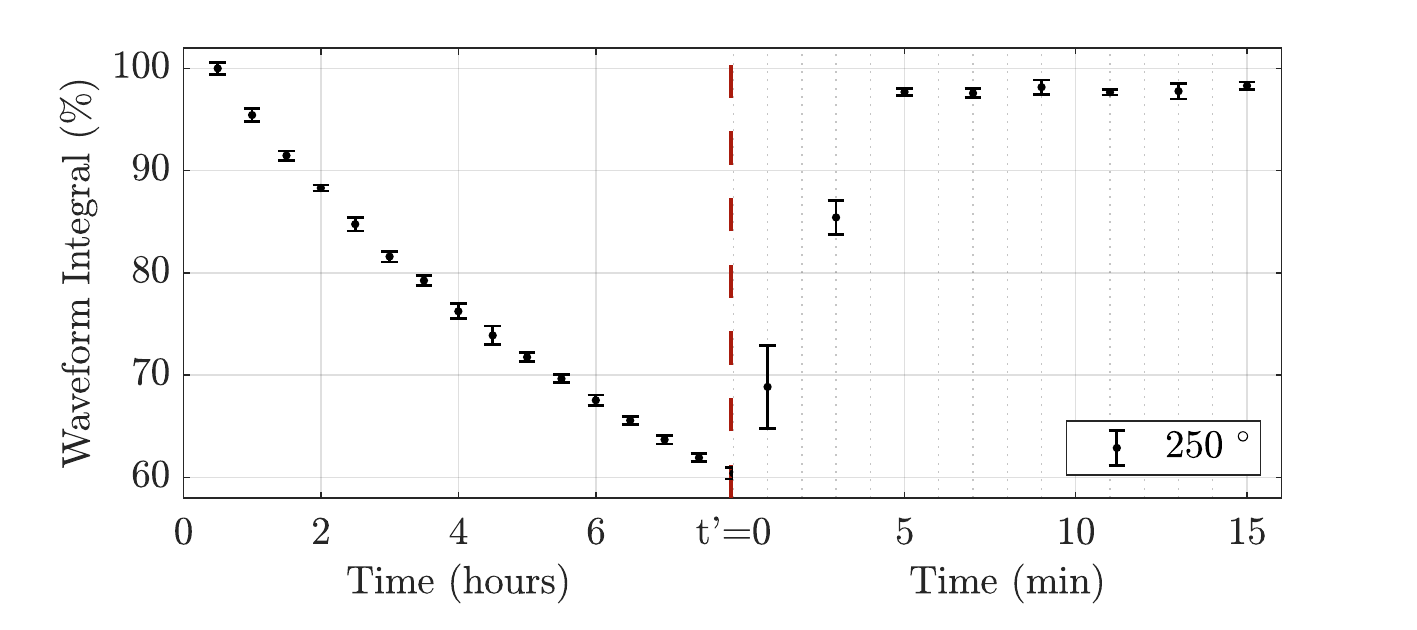}
\caption{\label{fig:time} EL average waveform integral as a function of the elapsed time after the gas circulation being stopped, being resumed at t=0.}
\end{figure}

As before, the Xe purity could be confirmed by a very good energy resolution achieved in the GPSC, about 8.0\% for 5.9 keV x-rays, and a fast recovery time of the scintillation, when gas circulation is resumed. In figure \ref{fig:time} the EL waveform integral is depicted as a function of the time elapsed after the Xe circulation was stopped, closing the getters' valves. In the absence of purification, the amplitude of the EL pulse slowly decreases due to the increase of impurities in the gas, as a consequence of outgassing from the surfaces inside the chamber. However, as soon as the gas circulation is resumed the EL average waveform integral recovers from 60 to 98\% within few minutesand, then to 100\% within 1.5 hours, denoting the efficiency of the purification system.

\subsection{Triplet lifetime, quenching and impurity concentrations}
\label{sec:quenching}

When energetic charged particles cross a volume of gaseous xenon, they either excite or ionize the atoms. Three-body collisions between one excited or ionized atom and two ground state atoms produce excimers. The excimers are produced in either singlet, $(^{1} \sum _u ^+)$ or  $(^3P_1)$, or triplet, $(^3\sum _u^+)$ or $(^3 P_2)$, molecular states, with lifetimes of about 4.5 ns and 100 ns, respectively, which decay to the unbound ground state emitting scintillation.  

In the presence of impurities, it is possible that xenon excimers undergo through quenching, non-radiative collision with molecular impurities, in competition with the scintillation emission process, reducing the scintillation output. Due to its very fast decay time, the singlet state is not strongly affected by this process. However, even low impurity concentrations can significantly decrease the triplet lifetime, which is even more sensitive to impurities than the scintillation yield itself. 

Although it is impossible to measure the singlet lifetime in our system, due to time response limitations of our PMT, the effective time decay constant of the Xe triplet state ($\tau_3^*$ ) can be measured from the average primary scintillation waveform obtained from $\alpha$-particle interactions. For this purpose, the detector was radiated with $~$2.5-MeV $\alpha$-particles produced by a Am-241 radioactive source. A Mylar film was placed between the radioactive source and the detector window to degrade the energy of the $\alpha$-particles. The reduced electric field in the drift region was set to 140 V cm$^{-1}$bar$^{-1}$ to avoid ion recombination and the electric field in the EL region was set to zero, since the secondary scintillation signals are not required. 

The S1 pulses from $\alpha$-particle interactions are large enough to be detected in individual waveforms. An average waveform was computed from about 10³ primary scintillation events, after applying the pre-processing algorithm and corrected each individual waveform for the baseline offset, measured before the S1 pulse. In addition, before averaging, the individual waveforms were aligned using the time corresponding to 50\% of the rising edge of the S1 pulse.

The triplet decay component of pure Xe is about 100 ns, which is reasonably faster than the PMT afterpulsing. Therefore, the PMT afterpulsing is distinguishable from the primary scintillation pulses and we can select a time-window where it does not yet influence the S1 pulse. 

\begin{figure}[ht]
\centering 
\includegraphics[width=0.8\textwidth,clip]{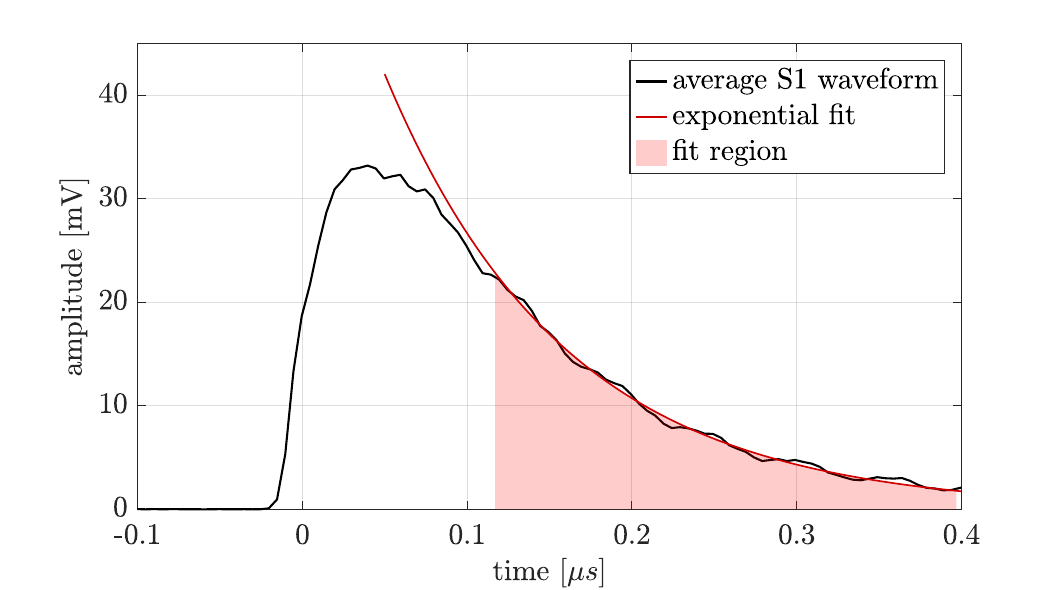}
\caption{\label{fig:triplet} Average waveform for the primary scintillation pulse produced by 2.5 MeV $\alpha$-particle events, including the exponential fit to the tail.} 
\end{figure}

An exponential function is fitted to the tail of the average primary scintillation pulse, figure \ref{fig:triplet}, between 120 ns and 400 ns, to avoid both the Xe fast scintillation component (singlet) and the PMT afterpulsing. In this way, a value of $109.4 \pm 8.2$ ns (95\% C.L.),  which agrees within the experimental error with the mean value measured in pure Xe, $\tau_3=100.9 \pm 1.4$ ns (95\% C.L.) \cite{moutard}. This value implies that the quenching effect is negligible to be noticed in the triplet decay time, for the impurity content in our gas. Nevertheless, assuming the worst-case scenario, i.e. 101.2 ns for the obtained experimental triplet decay time and 102.3 ns for the reference value, the lower and upper values for 95\% C.L. uncertainty, respectively,  maximum impurity concentrations may be estimated.

Knowing the effective lifetime of the triplet, maximum impurity concentrations ($f$) are estimated from the two-body quenching rates ($K_{2}$), reported in the literature for the major impurity species, through equation \eqref{eq:lifetime} \cite{microscopic}, 

\begin{equation}
\label{eq:lifetime}
\tau_{3}^{*} = \frac{\tau_{3}}{1+f k_{2} \tau_{3}},
\end{equation}
yielding to an upper limit of about 4 ppm for the impurity concentration  (CO$_2$, O$_2$, H$_2$O, and CH$_4$), and 82 ppm for N$_2$. Table \ref{tab:quenching} summarizes these results.

\begin{table}[ht]
\centering
\caption{\label{tab:quenching} Maximum impurity concentrations calculated from equation \eqref{eq:lifetime}. }
\begin{tabular}{|l|c|c|}
\hline
Impurity& K$_{2}$ (cm$^3$ s$^-1$) at 1.24 bar & $f$ (ppm) \\
\hline
N$_{2}$ & 0.569 \cite{velazco} & <82\\
CO$_{2}$ & 13.5 \cite{velazco} & <3.5\\
O$_{2}$ & 13.2 \cite{velazco} & <3.5\\
H$_{2}$O  & 11.2 \cite{balamuta} & <4.2\\
CH$_{4}$ & 14.7 \cite{velazco} & <3.2\\
\hline
\end{tabular}
\end{table}

\subsection{Electron lifetime and attachment}
\label{sec:attachment}

\begin{figure}[ht]
\centering 
\includegraphics[width=0.8\textwidth,clip]{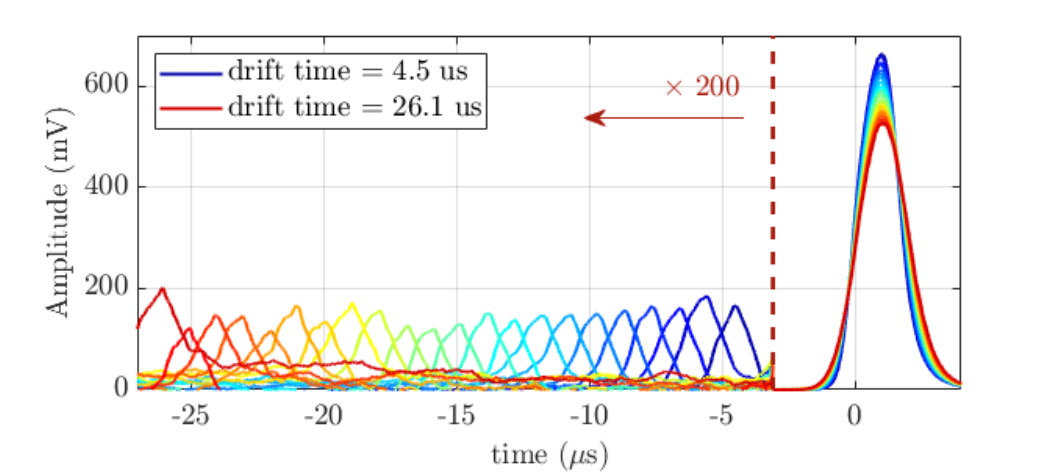}
\caption{\label{fig:penetration} Primary and secondary scintillation average waveforms for different 22.1 keV x-ray interaction depths, defined by the drift time of the primary electron cloud.}
\end{figure}

\begin{figure}[ht]
\centering %
\begin{center}
\includegraphics[width=1\textwidth]{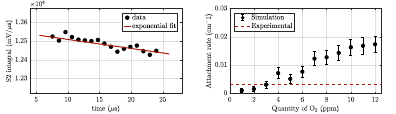}
\caption{\label{fig:lifetime} (Left) Integral of the secondary scintillation average waveform as a function of the drift time of the respective electron cloud, including an exponential fit to the data. (Right) Simulation results obtained for the attachment rate of O$_{2}$ as a function of its impurity content in pure Xe.}
\end{center}
\end{figure}

To measure the electron lifetime in our chamber, the average time an electron drift through the xenon before being captured by impurities, due to attachment, we resort to EL signals, S2. Since the S2 yield is proportional to the number of electrons crossing the scintillation region, any loss of electrons during the primary electron cloud drift in the absorption region results in a smaller S2 signal. The GPSC was irradiated with 22.1 keV x-rays from a $^{109}Cd$ source and we selected different sets of waveforms corresponding to different x-ray interaction depths, which were obtained from the drift time of the primary electron cloud, the time difference between S1 and S2 pulse rising edges in individual waveforms. For that, only waveforms with a distinguishable S1 peak were used. An average waveform was obtained for each set, figure \ref{fig:penetration}. For each average waveform, the amplitude of the S2 signal was obtained by integrating the region between 5\% of the rising and falling edge. By plotting the S2 amplitude as a function of the drift time and fitting this data to an exponential curve, the electron lifetime is obtained from the decay constant of the exponential fit, figure \ref{fig:lifetime}. 

The electron lifetime was measured for different electric field intensities in both the absorption and EL regions, and a value of  2.1 $\pm$ 0.1 ms was obtained, which is in good agreement with the values reported in the literature for state-of-the-art experiments \cite{martinez, njoya}. The secondary scintillation yield values obtained from S2 waveform integration are in good agreement with those obtained in a driftless GPSC \cite{monteiro}, supporting the reliability of the average waveform analysis. In addition, longer path lengths drifted by the electrons will result in larger electron clouds, due to diffusion, producing longer S2 pulses, a behaviour that is also reflected in the obtained S2 waveforms, figure \ref{fig:penetration}.

Knowing the electron lifetime and the drift velocity, $v_d$, computed from the drift time of the x-ray events taking place close to the radiation window, we can assess the attachment coefficient ($\eta$) of the gas in our chamber using equation \eqref{eq:attachment} \cite{gomez_cadenas}. An attachment rate of 0.003 $\pm$ 0.001 cm$^{-1}$ was, then, estimated in our chamber.

\begin{equation}
\label{eq:attachment}
    \eta = \frac{1}{\tau_{lifetime}}  \frac{1}{v_{d}}
\end{equation}

Attachment rates were simulated, using state-of-the-art electron transport tools (Magboltz), for the most common electronegative impurities (e.g., H$_{2}$O, N$_{2}$, O$_{2}$, CO$_{2}$). Operating parameters, such as pressure, temperature, and drift electric field, from the experimental conditions were taken into account in the simulation.  Magboltz was used to determine the attachment rate for the different impurities as a function of the respective concentrations in pure Xe. 

The simulation results for O$_{2}$, which is a highly electronegative molecule, show that attachment rates of 0.003 cm$^{-1}$ are only possible for impurity concentrations below 3 ppm, see figure \ref{fig:lifetime}-right, in line with the result obtained from the quenching rates in section \ref{sec:quenching}. For the other molecules, no conclusions can be drawn from the obtained attachment rate, since such value is compatible with impurity concentrations much higher than the upper limits obtained from the quenching data, being the attachment negligible for the maximum concentration values presented in Table \ref{tab:quenching}.

\section{Conclusions}
\label{sec:conclusions} 

Up to now, the effectiveness of the gas purification technique based on heating tapes rolled up around a vertical tube, in the outside, to heat SAES St707 getters placed inside the tube and promoting gas circulation by convection has only been confirmed by the state-of-art energy resolutions achieved in our detectors and by the fast recovery of the scintillation output when the gas circulation is resumed after being stopped for some time. 

Further evidences are now produced by analysing the waveforms of primary and secondary scintillation pulses. In the primary scintillation pulses, the decay time of the triplet state of the Xe excimer was measured to be $100.9 \pm 8.2$ ns (95\% C.L.). From this value and from the two-body quenching rates reported in the literature, an overall maximum impurity concentration below 4 ppm can be estimated for molecular species such as H$_{2}$O, O$_{2}$, CO$_{2}$, CH$_{4}$ and other heavier molecules, and below 82 ppm for N$_{2}$. The electron lifetime in our chamber was found to be 2.1 $\pm$ 0.1 ms, indicating an electron loss below 1 $\%$ during the drift across the full, 3.6-cm absorption region, a value that is similar to those obtained in high-performance optical TPCs.

\acknowledgments

This work was fully funded by national funds, through FCT - Fundação para a Ciência e a Tecnologia, I.P., under Projects No. UIDP/04559/2020, UIDB/04559/2020, CERN/FIS-TEC/0038/2021 and PTDC/FIS-NUC/3933/2021, and under Grant No. UI/BD/151005/2021.

\end{document}